\documentclass[pra,twocolumn,nofootinbib,eqsecnum,floatfix]{revtex4}
\usepackage{bm} % bold math
\usepackage{amsmath}

\def\eq#1{eq.$\,$(\ref{#1})}

\begin{document}

\title{Are We Typical?}

\author{James B. Hartle}
\email{hartle@physics.ucsb.edu}

\author{Mark Srednicki}
\email{mark@physics.ucsb.edu}

\affiliation{Department of Physics,
 University of California,
 Santa Barbara, CA 93106-9530}

\date{May 2, 2007}

\begin{abstract}

Bayesian probability theory is used to analyze the oft-made assumption that humans are typical observers in the universe.  Some theoretical calculations make the {\it selection fallacy\/} that we are randomly chosen from a class of objects by some physical process, despite the absence of any evidence for such a process, or any observational evidence favoring our typicality.  It is possible to favor theories in which we are typical by appropriately choosing their  prior probabilities, but such assumptions should be made explicit to avoid confusion. 

\end{abstract}

%\pacs{PACS }

\maketitle

An increasingly common kind of reasoning in fundamental cosmology starts from an assumption
that some property of human observers is typical in some class $\cal C$ of objects in the universe. 
This assumption is then used in conjunction with a physical theory of cosmology to draw conclusions about the properties of the universe that we should expect to observe. However,  it is perfectly possible 
(and not necessarily unlikely) for us to live in a universe in which we are not typical.  
Indeed, there are many choices of the class $\cal C$ (such as all compact objects with mass between 10 and 100$\,$kg) for which we have data that definitively establishes that humans are not typical.  Any assumption of typicality or mediocrity should therefore be carefully scrutinized. 

Such an examination seems especially timely because of recent discussions drawing conclusions from the possible existence (in a very large universe) of vacuum fluctuated brains (also known as ``Boltzmann brains'' or ``freak observers''). Astounding conclusions have been drawn using arguments that involve these possibilities, notably those of Page \cite{Page06a,Page06b,Page06c,Page06d} that the universe must be limited in time and space 
in order to avoid having too many fluctuated brains that would make 
human observers untypical in any class that includes these fluctuated brains.  
Some other recent discussions of this issue include \cite{DKS02,BF06,Linde06,Vil06,Banks07}.  
Early discussions of the role of typicality in inflationary cosmology include \cite{BLL94,Vil95}.
Previous criticism of typicality assumptions includes \cite{Smo04,Neal06}.

Typicality assumptions involving human observers in a class containing other kinds of observers can be based on different types of information.  These include
(1) empirical extrapolations of data concerning extraterrestrial observers, 
(2) plausible guesses for the predictions of a more fundamental theory, and 
(3) mere personal preference for theories in which we are typical of something. 

Since we currently have no data on extraterrestrial observers, 
there are no examples of type (1). 
An example of type (2) is the oft-made assumption in multiverse cosmology  that the number of ordinary observers is proportional to the number of galaxies, or to the number of baryons, or to some other similar proxy.  This paper does not criticize such assumptions when they are clearly stated. However, it will be argued that assumptions of type (3) are not well justified, and can lead
to ridiculous conclusions that will be illustrated with simple examples. 

To indicate where the paper
is headed some of its conclusions are summarized below.

\begin{itemize}

\item A theory is not incorrect merely because it predicts that we are atypical.

\item Theories are tested using our data as a collectivity of human observers. What other observers might see, how many of them there are, and what properties they do or do not share with us are irrelevant for this process. 

\item No part of our data should be neglected in the process of discriminating between competing theories unless it can be demonstrated that the relevant probabilities are insensitive to it. 

\item Two theories that predict our data with equal probability are indistinguishable on the basis of that data. Cosmological models that predict that at least one instance of our data exists (with probability one) somewhere in spacetime are indistinguishable no matter how many other exact copies of these data exist. 

\item We have data that we exist in the universe, but we have no evidence that we have been selected by some random process. We should not calculate as though we were.  

\item In a fundamental theory of quantum cosmology, there is no need for  any assumption of typicality to predict what we might see. However typicality, and the measures to implement it, may be implied by the fundamental theory. In absence of a manageable candidate for such a theory, it can be useful to try to guess such measures. 

\end{itemize}

Bayesian probability theory is the basis for this analysis.  
In addition to supporting the conclusions above, it will be used  discuss some classes of mistakes that can be made in applying the Bayesian procedure. These include neglecting relevent parts of our data, and making the {\em selection fallacy} that we are randomly selected from some class $\cal C$. 

\section{Testing theories using Bayesian Probabilities}

This section briefly reviews the elements of the Bayesian framework for testing theories. 
(For a compact introduction,
see \cite{Sred05}; for complete details, see \cite{Jaynes}; for an elegant presentation of the 
necessity of the Bayesian point of view, see \cite{App04}.)  Suppose we have a set of theories that we wish to discriminate between.  For simplicity, assume the theories  form a finite discrete set
$\{T_1,  \ldots, T_N\}$, with any continuous parameters binned.   The specification of a theory is taken to include both the dynamics and the initial conditions, so that the theory contains enough information to make calculable predictions.   For simplicity, we assume that the set of theories
is complete, so that one of them must be true; if this is not the case,
the probabilities to be calculated are relative rather than absolute.  

Bayesian analysis has three inputs:  First is a {\it prior probability\/} or {\it prior\/}
$P(T_i)$ for each theory.  These priors should be assigned before any data is obtained.  
Although the priors can in principle be chosen arbitrarily, standard practice is to try to make 
a choice that  suitably reflects our ignorance; for example, by assigning equal probabilities to the discrete set of theories. In the terminology of Bayesian theory, the choice should be ``noninformative''.  
The exact choice of priors is ultimately unimportant in situations where there can be an arbitrarily large number of identical trials. However, in areas like cosmology, where data is hard to come by, priors can strongly influence the final conclusions.  

Second there  is the specification of the data $D$ with which we seek to discriminate between the theories. This is the data of the collectivity of human observers doing the Bayesian analysis. 

The third input  is the {\it likelihood\/} $P(D|T_i)$ that (upon suitable observation) we will
obtain a particular data set $D$, given that the theory is $T_i$.  
An important point is that, for each theory, the likelihood
should be {\it calculable}.  If we can only estimate but not calculate a likelihood, then we should enlarge our set of theories by introducing new parameters that account for our ignorance of the
likelihood, and allow these parameters to vary over the new, larger, set of theories.  

Finally, there is the output --- what we really want to know --- the {\it posterior probability} $P(T_i|D)$, the
probability of theory $T_i$, given the data $D$ that we have obtained.  The posterior
probabilities for the various theories are related to the prior probabilities and the likelihoods
by {\it Bayes' theorem},
\begin{equation}
P(T_i|D) = \frac{P(D|T_i) P(T_i)}{\sum_i  P(D|T_i) P(T_i)}  .
\label{bayesthm}
\end{equation} 
This is the key formula of Bayesian analysis. 

\section{Priors favoring Typicality} 

Our typicality can be enforced simply by choosing priors that favor theories that imply that we are 
typical. A specific example is useful to illustrate the dangers inherent in this kind of reasoning 
(type 3 above). 

Consider two theories of the development of planet-based intelligent life,
based on the appropriate physics, chemistry, biology, and ecology.
Theory $A$ predicts that there are likely to be intelligent beings living
in the atmosphere of Jupiter; theory $B$ predicts that there are no such beings.
Because Jupiter is much larger than the Earth, theory $A$ predicts that
there are today many more jovians than humans.  

Would we reject theory A solely because humans
would not then be typical of intelligent beings in our solar system? Would we use this theory to predict that there are no jovians, because that is the only way we could be typical?  
Such a conclusion seems absurd.

We could nevertheless enforce an assumption of typicality by assigning a low prior probability to 
theory $A$, and a high prior probability to theory $B$.  Alternatively, we could choose priors that favor theory $B$, so that the Earth is more typical of the planets in the solar system\footnote{
This was the point of view taken by  Huygens \cite{Huy1698} in 1698:
``A Man that is of Copernicus's Opinion, that this Earth of ours is a Planet, carry'd round and enlighten'd by the Sun, like the rest of them, cannot but sometimes have a fancy, that it is not improbable that the rest of the planets must have their Dress and Furniture, nay and their inhabitants too as well as this Earth of ours.''  Further he argues from the similarities among species from distant parts of the Earth (such as Europe and America) that we should be typical of the inhabitants of other planets: ``'Tis more probable that all the difference there is between us and them, springs from the greater or less distance and influence from that Fountain of Heat and Life the Sun; which will cause a difference not so much in their Form and Shape, as in their Matter and Contexture.''  He then concludes that there must be 
substances on the other planets serving the role of water on this one: ``I can't say that they are exactly of the same nature with our Water, but that they should be liquid their use requires, as their beauty does that they should be clear.''  In this case, Huygens' assumption of the typicality of the Earth obviously did not lead him to correct conclusions.}.  
However, any {\it a priori\/} favoring of either theory is contrary to the standard 
scientific practice of assigning noninformative priors that do not preselect conclusions before relevant data is obtained.

\section{Data}  
\label{Data}

A discussion of what is meant by `our data' necessarily begins with a discussion of what is meant by `we', `us', and `our'. In this paper, `we' refers to the information gathering and utilizing system (IGUS) that is engaged in the process of gathering data, constructing alternative theories, and using Bayesian analysis to discriminate between these theories. In this epoch, that is naturally the human scientific IGUS on Earth (HSI). If we ever have scientific exchanges with IGUSes 
elsewhere, the definition would have to be enlarged. 

 `Our data'  $D$ includes every scrap of information that the HSI possesses about the physical universe: every record of every experiment, every astronomical observation of distant galaxies, every available description of every leaf, etc., and necessarily every piece of information about the HSI itself, its members, and its history. The choices of coarse graining used to define this information, the level of accessibility demanded, etc., are inevitably subjective but assumed here to be fixed.  The data must all be expressed in physical terms in the language of the theories being tested so that the likelihoods $P(D|T_i)$ can be computed. 

All the information  possessed by the HSI should be included in $D$ unless it can be demonstrated that the posterior probabilities $P(T_i|D)$ are insensitive to it. Of course, we can practically test theories only if we can assume that this is the case for most of our data. The important point is that any such assumption can be checked by experiment or calculation. To arbitrarily neglect some pieces of data in favor of others is unscientific and risks contradiction when the neglected data is later considered. 

What are the implications of this for typicality?  It is {\it our data} that is used in a Bayesian analysis to discriminate between theories. What other hypothetical observers with data different from ours might see, how many of them there are, and what properties they might or might not share with us (defining some notion of typicality) are irrelevant for this process. In particular the number of vacuum fluctuated brains with different data (for example with disordered observations as discussed by Page \cite{Page06a, Page06b}) is irrelevant. As far as Bayes' theorem is concerned, the only other observers that matter are ones that have exactly the same data $D$.  

Observers with identical data may occur in multiple places in a large universe.  For example, in contemporary inflationary models, we may know that we are 14$\,$Gyr from the nucleation of our pocket universe. But we know neither the location of that pocket universe in spacetime or how many others like it with identical data may have been nucleated elsewhere. Vacuum or thermal fluctuations that produce identical data must also be considered. These unknown locations and origins must be summed over in calculating the likelihoods $P(D|T)$. {\it All we know is that there exists at least one such region containing our data.}  This idea is illustrated with simple models in the following sections. 

Beyond testing theories, questions sometimes arise like ``What would be the probability to observe a cosmic microwave background (CMB) at a temperature of 3$\,$K if we didn't already know the value of this temperature?''  In such cases the data $D$ (which includes the observation of a 3$\,$K CMB) must be divided into a part that could have varied and a notion of `us' that could not. There is additional ambiguity, and additional subjective choice, in making this division. At one extreme, the 3$\,$K could be included in the meaning of `us' (no division), in which case `we' could never have observed anything else, by definition. At the other extreme, `us' could be defined to be an object less than 200$\,$kg, in which case we could be talking about the CMB temperature a rock might have received.  Most physicists would make less extreme choices, but it is important to recognize that the answers to such questions depend on the choice. 

\section{Likelihoods and the Selection Falacy}

Return for the moment  to the human/jovian model and assume that each theory is able to predict a probability for the
number of humans $H$ and jovians $J$ alive today.  
The question then becomes, what is the data?  
We know that the actual number of humans today is ${H^*}=6.57\times 10^9$.  
(For simplicity, assume that this number is known exactly.)   
We know nothing at all about the actual number of jovians today; this data has not yet been 
obtained.  Thus, the likelihood that a theory $T_i$ predicts the data $D$ that we actually have 
in this case is given by summing over the mutually exclusive probabilities for different numbers
of jovians, with the number of humans fixed to its known value,
\begin{equation}
P(D|T_i) = \sum_{J=0}^\infty P({H^*},J|T_i)  .
\label{PDT}
\end{equation} 
The likelihood for the data is independent of the number of jovians predicted by
the theory, simply because we have not observed the number of jovians.

It is important to note that there is an alternative, incorrect computation of the likelihood that 
favors typicality.   The invalid reasoning goes as follows.  ``I observe that I am human.  
I might, however, have been jovian.
If there are $H$ humans and $J$ jovians, the probability that I observe myself to be human
is $H/(H+J)$.  Therefore, I should multiply the probability $P(H,J|T_i)$ that there are 
$H$ humans and $J$ jovians by a factor of
${H}/({H}+J)$ to account for the probability that I have observed myself to be human and not
jovian.  Then I compute the likelihood as
\begin{equation}
P(D|T_i) = \sum_{J=0}^\infty \frac{H^*}{H^*+J} P({H^*},J|T_i),
\label{PDT2}
\end{equation} 
instead of \eqref{PDT}.
The factor of ${H^*}/({H^*}+J)$ in \eqref{PDT2} results in likelihoods that 
(when used in Bayes' theorem) disfavor theories that predict $H \ll J$,
and hence lead to the conclusion that humans should be typical. This is the kind of argument made in \cite{DKS02,BF06}, and would seem to be a natural interpetation of
\cite{Page06a,Page06b,Page06c,Page06d}\footnote{However, in private discussions, 
Page says the typicality factors arise naturally in his theory of quantum mechanics \cite{Pageqm},
in which the basic quantities are operators representing the conscious perceptions of observers.  These ideas are not discussed here.}. 

The factor of $H^*/(H^*+J)$ would be correct if we (or some outside agency) 
had performed an experiment that had (somehow) selected a single intelligent being in the 
solar system, with equal {\it a priori\/}
probability for each, and that we or the outside agency  knew the result of this selection. 
However, noticing that we are human is not equivalent to performing such an experiment.  
For example,
it would certainly be incorrect to conclude that, since each of the two authors of this paper
is human, the correct factor to multiply by is $({H^*}/({H^*}+J))^2$.  This is because the 
probability that one of us is human is obviously not independent of the status of the other.
Thus, whatever selection is performed when one of us notices that he is human,
it is surely not equivalent to a random selection of a single intelligent being anywhere 
in the solar system.  

{\it  In fact, there has been no selection at all.  }
Intelligent beings on Earth notice that they exist, call themselves ``humans'', 
and count up how many there are.  Whatever may
or may not be happening on Jupiter is entirely irrelevant.
This leads immediately to \eq{PDT} for the likelihood, which does not 
favor or disfavor typicality of humans.  
If we now set all priors equal, $P(T_i)=1/N$ with
$N$ the number of theories, we immediately find $P(T_i|D) \propto P(D|T_i)$.
That is, the probability that a theory is correct is directly proportional to the probability
that it predicts the correct number of humans, independent of the number of jovians.

To compute likelihoods as though we had been randomly selected by some physical process, when there is no evidence for such a process, commits what might be called the
{\it selection fallacy.}  We are not a disembodied entity that was randomly selected to have a particular physical description; instead, we are the meaning of a transcription of `we' 
into the language of physical theory. 

As shown above, the correctly computed likelihood does not favor typicality of humans.
However, we can, if we wish, implement an assumption of typicality through the choice
of priors.  We do this by identifying the set of theories that predicts that $P(H,J|T_i)$ 
is small whenever $H$ is much less than $J$.   Then we assign these theories small values of $P(T_i)$.  For example, we could choose $P(T_i) \propto {\bar H}_i/({\bar H}_i+{\bar J}_i)$ 
(or some power of that), where $\bar H_i$ and $\bar J_i$ are the mean number of 
humans and jovians that are predicted by theory $T_i$. 

As was mentioned earlier, there is nothing wrong in principle with any choice of priors.  However,
choosing priors that favor one or another prediction (that we happen to like) goes against standard scientific practice, and simply represents an unsupported (by data) personal preference for a
certain result.   

\section{A Simple Cosmological Model}  

A simple (highly idealized) cosmological model will further illustrate the key issues. 

Consider a model universe which has $N$ cycles in time, $k=1, \ldots, N$. In each cycle the universe may have one of two global properties: red ($R$) or blue ($B$), which could be 
thought of as (for example) two different possible values of the CMB temperature.   To further simplify the discussion, the only relevant observables are assumed to be (1) the value of the property and (2) the existence of an observing system that is able 
to determine this value. 
In each cycle, the probability for such an observing system to exist is taken to be $p_E$; this probability is assumed to be independent of whether the universe is red or blue in that cycle. Furthermore, observations are assumed to be perfectly accurate, so that if red is observed in any cycle, then the universe is red in that cycle, and conversely.

Two competing theories of this model universe are proposed. One, {\it all red\/} or $AR$, 
in which all the cycles are red, and another, {\it some red\/} or $SR$, in which some number of particular cycles are red and the rest are blue. We (an idealized observing system) seek to discriminate between these two theories on the basis of our data. 

Suppose that we (a particular observing system) observe red. Our data $D$ is then $(E,R)$, 
which in the context of the model could be more fully described as `there is at least one cycle in which an observing system exists and the universe is red'.   Using this data we carry out a Bayesian analysis to discriminate between the two theories assuming equal priors for simplicity. The probability that there is at least one cycle with $(E,R)$ is the same as one minus the 
probability of the negation of this, which is 
the probability that no observing system exists in a cycle in which the universe is red. Since the probability for an observing system {\it not\/} to exist in any one cycle is $1-p_E$, the likelihoods are
\begin{equation}
P(E,R|T) = 1 - (1-p_{E})^{N_R(T)}  ,
\label{rblikelihoods}
\end{equation}
where $N_R(T)$ is the number of red cycles in theory $T$, equaling $N$ when $T$ is  $AR$. Assuming equal priors, \eqref{bayesthm} yields the posterior probabilities 
\begin{equation}
P(T|E,R) = \frac{P(E,R|T)}{P(E,R|AR) + P(E,R|SR)},
\label{rbposteriors}
\end{equation}
where $T$ is either $AR$ or $SR$. 

A number of limiting cases of the likelihoods \eqref{rblikelihoods} and their consequent posterior probabilities \eqref{rbposteriors} are of interest.   

If $p_E$ is close to one, then 
$P(E,R|T) \approx 1$ for {\it both\/} theories. That is not very surprising, since the probability is high for there to be an observing system in every cycle. The probability that there is at least one red cycle with an observing system is therefore also high in both theories. Thus our data do not discriminate between the two theories, and indeed  $p(T|E,R) \approx 1/2$ for both. 

Another case is where our data does not discriminate between the two theories occurs when 
$p_E <1$, and $N_R$ is large in {\it both} theories to make $(1-p_E)^{N_R(T)}\ll 1$, so that
$P(E,R|T)\approx 1$ for both theories. 
Even though there may be  more many more red cycles in the $AR$ theory than the $SR$ theory, the probability that there is {\it at least one} red cycle with an observing subsystem approaches one for both theories when $N_R$ becomes large in both. In this case, our data is not enough to discriminate between the two theories. 
This is important because inflationary theories can be expected to imply large numbers of any conceivable situation. 

If $p_E \ll 1/N$, then $P(E,R|T) \approx N_R(T) p_E$, and the resulting posterior probabilities are 
\begin{equation}
P(AR|E,R) = {1\over 1 + f_R},  \quad P(SR|E,R) = {f_R\over 1 + f_R},
\label{smpepost}
\end{equation}
where $f_R \equiv N_R(SR)/N$ is the fraction of red cycles in the $SR$ theory.  In this case,
if $f_R$ is small, the $AR$ theory is strongly favored.

In the small $p_E$ limit, the result \eqref{smpepost} is the same as it would be if it was assumed that we were the unique observing system existing in the universe, but that we did not know the cycle $k$ in which we were located.  More specifically,  were  we unique, the probability 
that we were in cycle $k$ observing $R$ would be $1/N$
if cycle $k$ is red in theory $T$, and zero if it is blue. Summing these probabilities 
over $k$ yields \eqref{smpepost}.

Confusingly, this result is also the same under the assumption that our location is randomly selected from the $N$ possible cycles.  This notion presupposes that we exist separately from our physical description.  But  we are not separate from our physical description in our data; 
we {\it are} the physical system described, and no random selection has been made.
In contemporary inflationary models, we are very unlikely to be unique as physical systems within the infinite multiverse.  In this case, reasoning as if we had been randomly selected will not give correct results, as the above example shows.  

The implications of this simple model for typicality can now be discussed.  To emphasize that other observers with data different from ours are irrelevant for predicting what we observe, the likelihoods \eqref{rblikelihoods} can be calculated in a different way.  First consider the probability $P(n_R,n_B|T)$  that there are $n_R$ observing systems in red cycles and $n_B$ orbserving systems in blue cycles,
\begin{align}
P(n_R,n_B|T) = &{N_R(T) \choose n_R}{N_B(T) \choose n_B} \nonumber \\
&\times p_E^{n_R+n_B} (1-p_{E})^{N-n_R-n_B},
\label{numbers}
\end{align}
where $N_B(T)$ and $N_R(T)$ are the number of blue and red cycles in each theory. 
Our data is `there is at least one red cycle containing an observing system'. This is related to \eqref{numbers} by 
\begin{equation}
P(E,R|T) = \sum\limits_{n_R=1}^{N_R(T)}\sum\limits_{n_B=0}^{N_B(T)}P(n_R,n_B|T),
\label{sumoverblue}
\end{equation}
which, as is easily shown, yields \eqref{rblikelihoods}. The elementary but important point is that the number of oberving systems measuring blue must be summed over. The value of that number therefore does not affect the probability for our data. 

If the number of red cycles is infinite in both models, then the two theories are not distinguished by our data even if the number of fraction of red cycles is small in $SR$. That is because our data is that there is {\it at least} one red cycle. As the number of red cycles approaches infinity, that becomes a certainty in both theories. Even though the typical observing system in the $SR$ theory is observing blue, our data provides no evidence that we are typical. 

In such cases it is tempting to reason as follows:  In the $SR$ theory with mostly blue cycles we are more likely to observe blue than red. Since we observe red we should reject this theory. This is the kind of argument made by \cite{DKS02} against eternal deSitter space.  This, however, is an instance of the selection fallacy, and in more familiar situations leads to absurd conclusions such as those discussed earlier. 

\section{Deriving Measures}
 After a sufficient time,  an isolated quantum system with (sufficiently complex interactions) will come to equilibrium, and the predictions for the probabilities of certain variables like total energy, momentum and number will be given by an equilibrium density matrix. In principle the there is no need for  an equilibrium density matrix. Were the initial state known and powers of computation unlimited, the state could be evolved forward by the appropriate time and the probabilities calculated directly. However,  the power, utility, and perhaps even the necessity of a notion of equilibrium and the density matrix that defines it are well known. The important point is that such a density matrix cannot be posited arbitrarily; it follows from the dynamics in principle, and its form can be checked (see, e.g., \cite{Sred94}). 

Similarly, in a complete quantum theory of the universe, there is in principle no need for any further assumption about typicality of human observers, or a measure implementing that notion of typicality.  The probabilities for our observations would follow from a theory of the dynamics, the initial quantum state, and an accurate quantum description of human beings, their knowledge, and their data \cite{Har04}.

However, we presently lack even one example of such a complete theory, and in practice
calculations of the history and observations of human observers are well beyond current 
and foreseeable abilities to compute.  
It then becomes useful to make plausible hypotheses of what such a complete theory might predict, and (as  mentioned above) to expand the family of theories being tested to include these hypotheses.  For example, inflationary cosmology frequently assumes a {\it classical} spacetime that models distributions of bubbles nucleated in quantum events. Our location in this spacetime is unknown. To calculate the probabilities of what we might see, it becomes convenient to posit a distribution of positions in space and time together with  a measure to define it.  This distribution can in principle be checked in a complete theory of quantum cosmology. It might even be possible to infer a distribution for where other observers are located in the universe and a measure to define that. 

In a Bayesian analysis of such situations, priors must be assigned to candidates for a complete theory, and also to different hypotheses of what it will predict. 
In the example of inflationary cosmology, priors would be assigned to different measures and typicality assumptions. 
In Bayesian terms, debates about which measure is correct are discussions of these priors.
The important thing is that these priors can be improved through the usual process of Bayesian updating after acquiring more data or by  firming up predictions by better calculation. That is in contrast to priors which merely express personal preference for theories in which we are typical. 

\section{Conclusion}  The course of physics, both theoretically and experimentally, is guided by prejudice as to the nature of the theory sought for. We favor theories that are simple, beautiful, precisely formulable mathematically, economical in their assumptions, comprehensive, unifying, explanatory, accessible to existing intuition, etc. etc. Most importantly we favor theories that are successful in predicting new data beyond what we have at the moment. The bases for such prejudices do not lie in logic but rather previous experience with constructing successful theories. 

Especially in areas far from immediate experiment, it is important to distinguish facts,  logical deduction, and prejudices. Bayesian analysis provides a framework for doing this.  Data are the domain of facts, likelihoods are the domain of logical deduction, and the priors are the domain of theoretical prejudice. 

This paper has analyzed typicality assumptions using the Bayesian framework. The main conclusions have already been given in the Introduction and we will not repeat them here.  At present, there are no observational data supporting an assumption that we are typical in some class of observers, and our understanding of biological evolution is insufficient to supply a theoretical justification through the likelihoods. Many calculations that produce likelihoods that favor our typicality do so via the selection fallacy; this can lead to absurd conclusions, as the human/jovian model shows.  

On the other hand, a preference for typicality can legitimately be made through a suitable choice of priors. However, such choices should be made explicit, so that others can properly evaluate the conclusions that are ultimately reached. 

\begin{acknowledgments}

We are grateful to Don Page for voluminous and clarifying communications, and to 
Tom Banks, Andre Linde, Leonard Susskind, Alex Vilenkin, and colleagues in Santa Barbara  for useful remarks and helpful criticism.   
This work was supported
by the National Science Foundation under grants PHY04-56556 and PHY05-55669.

\end{acknowledgments}

\end{document}